\documentclass{article}
\usepackage{arxiv}
\usepackage[utf8]{inputenc} 
\usepackage[T1]{fontenc}
\pdfoutput=1 
\usepackage{hyperref}       
\usepackage{url}            
\usepackage{booktabs}       
\usepackage{amsfonts}       
\usepackage{nicefrac}       
\usepackage{microtype}      
\usepackage{lipsum}
\usepackage{makecell}
\usepackage{hyperref}
\usepackage{float}
\usepackage{fourier} 
\usepackage{array}
\usepackage{graphicx}
\usepackage{tabularx}
\usepackage{appendix}
\usepackage{url}
\usepackage[caption=false,font=footnotesize]{subfig}

\newcommand*{\textcite}[1]{\citeauthor{#1}~\shortcite{#1}}
\newcommand*{\parencite}[1]{\cite{#1}}

\title{Does the First Mover Advantage Exist on GitHub?}

\author{
  Aditya Mehta, Arun Paudyal, Atul Sharma, Zyanya Ambros \\
  Institute for Web Science and Technologies (WeST)\\
  University of Koblenz-Landau\\
  Koblenz 56070, Germany \\
  \path{adityamehta, apaudyal, atulsharma, zambros@uni-koblenz.de} \\
   \And
 Ipek Baris, Jun Sun, Oul Han, Akram Sadat Hosseini \\
  Institute for Web Science and Technologies (WeST)\\
  University of Koblenz-Landau\\
  Koblenz 56070, Germany \\
  \path{ibaris, junsun, han, sadathosseini@uni-koblenz.de} \\
}

\begin{document}
\maketitle
\begin{abstract}
Collaborative consensus-finding is an integral element of many Web services and greatly determines the quality of information, content, and products that are available through the Web. That also means that the dynamics of democratic consensus-finding strengthen collective resilience against potential threats that attempt to degrade information, content, and products and affect Web data, users, behaviors, and even beyond as well as offline life. Even on Web platforms that are open to all, the influence of some first mover authors may shape future discussion and collaboration, which is comparable to academic citation networks for instance. In a social coding network such as GitHub, activities of a set of users can have influence on other users who can get interested in further actions, possibly contributing to a new project together with influential users. In this paper, we analyze the effect of contribution activities on gaining influence in this and comparable networks that provide users the functionality and aims for reaching collaborative goals on the Web. For this purpose, we present an empirical approach to identify the top influential users by using network features and contribution characteristics, which we find in existing and newly collected data set. We find that early adopter dynamics exist in the GitHub community, where early adopters have more followers in the end as expected. However, we also see counterexamples that arise due to the social networking behavior of late adopters, and due to the aging effect of older repositories and users. We publicly share the source code and the data sets for reproducing our paper.  
\end{abstract}


\section{Introduction}\label{introduction}

By testing an established concept of network analysis in the setting of discussion threads on GitHub, we utilize a familiar scenario in order to contribute to an immanent problem of web governance in the age of AI: To enable social, democratic, and equal deliberation among users while preventing clearly harmful statements or behaviors. Thus, by observing how constructive and solution-seeking opinions are exchanged among users towards the aim of finding consensus, and by evaluating which users gain influence over time over other users, we can get closer to best practices of democratic discussions and deliberative processes on social networks. We focus on one specific dynamic among many: Similar to a firm that gains a competitive advantage over its competitors by entering first in the market~\cite{suarez2005half}, early discussants may shape future discussions and outcomes on a platform with lasting imprint for the future. This is called the \textit{first mover advantage}. Its dynamics may influence \textit{deliberation}: the process of discussing until reaching mutually improved opinions \cite{wright2007democracy}.

As all git-based systems, Github\footnote{www.github.com} enables multiple users to work collectively and collaboratively on projects, which provides a process of consensus-finding with the aim of eliminating possible flaws and improving the quality of output.
In addition, GitHub integrates social network features such as following or starring, which makes it popular among users~\parencite{10.1145/2556420.2556483}. Activities by users known as ``influencers'' can affect the subsequent actions of other users due to closeness to project owner, long-term interaction, status, popularity, value, or participation~\cite{farias2018characterizes}. 

The above-mentioned manners of user interaction already induce a positive feedback loop: Users with high influence are more likely to accumulate influence later on. The influence of users, as measured by the number of followers for instance, can thus be described by the preferential attachment mechanism~\cite{albert2002statistical} or the \textit{rich-gets-richer} effect, which in this case would naturally favor older users (i.e., first movers) since they have more time to accumulate influence. However, since information technologies are often rapidly changing, repositories and users can suffer greatly from the \textit{aging effect}  \cite{medo2011temporal} and quickly lose relevance. This reflects the natural preference for newness. In most real systems, the two effects rich-gets-richer and aging coexist and interact.
For GitHub, which of the two is more effective determines whether the first mover advantage exists, and thus determines the deliberative quality and equality of the communities. In the extreme case of deliberative imbalance, the rich-gets-richer effect becomes so dominant that eventually the \textit{winner-takes-all} effect emerges~\cite{bianconi2001bose}, discouraging new users from contributing; or the aging effect becomes so dominant that all contributions get quickly forgotten. 

For ideal democratic deliberation, however, every claim should be treated equally and has the same chances to be considered~\cite{friess2015systematic}. In this sense, studying influential users on GitHub gives insight into participation, contribution and influence~\cite{badashian2014involvement} in consensus-seeking deliberative processes via online social networks.

Thus, we focus on the following four, fine grained, research questions in this study.
\begin{itemize}
    \item \textbf{RQ1:} Which metrics are appropriate to measure users' influence?
    \item \textbf{RQ2:} Do the users who joined GitHub early have the first mover advantage by amassing more followers and influencing others in the network?
    \item \textbf{RQ3:} Do users with higher follower counts make higher contributions on GitHub?
    \item \textbf{RQ4:} Can macroscopic observations of the follower network tell whether the first mover advantage exists?
\end{itemize}

Existing studies have explained the user influence and its dynamics on GitHub to some extent, however some of the research gaps still exist~\cite{bana2018influence}. To systematically tackle these research questions, we present an exploratory analysis to compare different metrics including a) the number of followers \cite{Badashian2016millionfollower, liao2017devrank, kobayakawa2019study}, b) HITS~\cite{kobayakawa2019study} and PageRank~\cite{bana2018influence} values, and c) user activity metrics such as pull-requests and issues~\cite{goggins2014connecting}. We also conduct observations at the macroscopic level, such as the in-degree distribution and the growth pattern of the follower network over the years, and compare the result with other networks such as academic citation networks as well as analytical models, where the first mover advantage may or may not exist.

The rest of the paper is organized as follows. In Section \ref{relatedwork}, we present an overview of existing studies related to the dynamics of influential users on GitHub. In Section~\ref{methodology} we briefly explain our methodology for the analysis. In Section \ref{results}, we present the dataset we have used, along with our technical experiments and the corresponding results. 
Finally, in Section \ref{conclusion}, we conclude with some discussions and point to potential improvements and future work. The source code and the dataset are publicly available here\footnote{\url{https://github.com/Institute-Web-Science-and-Technologies/First-Mover-Advantage-on-GitHub}}.

\section{Related Work}\label{relatedwork}
The availability of network of millions of repositories and users, and several features have made GitHub a promising platform to conduct several studies and researches~\cite{gousios2017mining}. Some of those studies are focused in understanding the dynamics of influential users and repositories on GitHub. Studies based on surveys \cite{blincoe2016understanding, farias2018characterizes} have attempted to determine the factors that make a user influential on GitHub. Though surveys could trigger possible bias in the conclusions, these studies have nonetheless revealed some traits such as popularity, contribution, value-generation, interaction as factors determining the influence of a user. These traits could be used as metrics for technical and comprehensive studies.

Representing content in the Web as networks has already shown great success in the field of Web Science~\cite{broder2000graph}. Typically, social entities such as individuals and organizations are modelled as nodes in the network, while relations between entities such as friendship and communications are modelled as links. Network models enable us to focus on the structural aspects of the Web, and benefit from the abundant tools available in network science and related fields, such as node centrality measures~\cite{das2018study} and preferential attachment models~\cite{kunegis2013preferential}.

Node centrality measures can be applied to understand the influential behaviour of users. For example, HITS \cite{kobayakawa2019study} and PageRank \cite{bana2018influence} are usually the standard to identify influential nodes in a network. In one of such studies, a custom metric called ``DevRank'' has been devised to mine the influential users based upon their followers count and code commit \cite{liao2017devrank}. A similar study identified the ``influence index'' for developers, repositories, technologies and programming languages \cite{bana2018influence}. This study was more comprehensive and was based upon the network analysis of repositories, developers and their followers. Their approach included multiple unexplored metrics like forks and watchers, and also overcomes traditional influence determination based upon followers count.

Preferential attachment refers to the mechanism that nodes have the preference to attach to other influential nodes in the network. It has been used to describe the growth of many systems on the Web~\cite{newman2001clustering,capocci2006preferential}. Preferential attachment models bridge this microscopic mechanism and many macroscopic behavior of networks. For instance, the very basic approach, the Barab\'{a}si-Albert model \cite{albert2002statistical}, predicts that the network has a power-law (scale-free) degree distribution~\cite{barabasi1999mean}. Further generalizations of the model including heterogeneous node fitness~\cite{bianconi2001competition} and aging~\cite{medo2011temporal} can produce broader degree distributions which are found in many real networks.

The ``first mover advantage'' refers to the phenomenon that a firm will have a competitive advantage over its competitors by entering early in the market \cite{suarez2005half}. Though this phenomenon is a novel topic to explore in GitHub, similar works have been done for other networks. Newman~\cite{newman2009first} has suggested that, earlier academic publications in each field gain more citations than late comers. In other words, the first mover advantage exists at the meso-level in citation networks.
However, further studies show large deviations in different scenarios. For instance, Sun et al. \cite{sun2020timeinvariant} have observed that papers published in different time period have strikingly similar citation growth patterns, suggesting that the first mover advantage is absent at the macroscopic level in citation networks due to the aging effect~\cite{medo2011temporal}. Adamic and Huberman~\cite{adamic2000power} have shown that the first mover advantage is absent in the World Wide Web.

\section{Methodology}
\label{methodology}
To systematically approach the research questions, we conduct a node centrality based analysis, a user activity based analysis and a macroscopic analysis of GitHub.

\subsection{Node centrality based metrics}
The number of followers of a user is a primary indicator of his/her popularity on the platform~\cite{blincoe2016understanding}. Often termed as ``degree centrality" \cite{Badashian2016millionfollower}, this behaviour is commonly exhibited by the users in social networks \cite{liao2017devrank, song2007identifying}. Following a user on GitHub helps the follower to be updated with the activities of the followee (commits, pull requests, comments). Therefore, a user with a large number of followers has great coverage since their activities are massively observed. However, rather than relying only on follower count, the better approach is to identify influence by HITS \cite{liao2017devrank} and PageRank algorithms \cite{bana2018influence}.

The HITS (Hyperlink-Induced Topic Search) algorithm~\cite{agosti2005theoretical} was originally devised to rate web pages, and has been applied to find central nodes in a network. This algorithm rates a node based on its hub score and authority score.
As stated by Manning et al. \cite{manning2009introduction} "A good hub page is one that points to many good authorities; a good authority page is one that is pointed to by many good hub pages". In network terms, the greater the number of hubs pointing at a node, the higher its authority score. We have applied this notion to find the influential user based on its authority score. 
Similarly, PageRank~\cite{pretto2002theoretical} is an algorithm based on random walk to find the relevance of a website on the Internet. The PageRank algorithm can be extended to find the relevance of nodes in a directed network. In this context, we also use it to measure the influence of a user on GitHub.

\subsection{User activity based metrics}
User participation by code contribution and engagement in discussions is typically observed in software social networks. Influence of users can be determined by quantifying such activities \cite{goggins2014connecting}. A typical user contribution activities in GitHub would be to create a pull request, to get it merged, to submit an issue \cite{goggins2014connecting}. Whenever a user wants to submit his code to the main repository, the user should open a pull request. The accepted pull requests cause merging of proposed code changes into the main repository. The number of pull requests and submission of issues can be considered as influential activities \cite{goggins2014connecting}. The inclusion of these user activity metrics will explain the influential behaviour of users from the software perspective. The network-based analysis will be further incorporated to explain the dynamics of the first mover advantage.

\subsubsection{Pull request}
User can submit a pull request to propose changes in a repository on GitHub. Other users of the project take a decision either to merge the pull request or to close it with comments. The merged requests add new code in the original repository \cite{tsay2014let}. A merged pull request is an indicator that a participant is influencing the project because other participants accepted their contribution as worthy or useful \cite{goggins2014connecting}.

\subsubsection{Issues}
Issues in the repository are offered to report software bugs, to bring enhancements, collect user feedback, and organize tasks. Issues can act as more than just a place to report software bugs since user influence can be related to the submission of an issue \cite{goggins2014connecting}.

\subsection{Observing macro-level behavior}
In addition, we can gain insights into GitHub by observing the macroscopic behavior of its user interaction network, in particular, its degree distribution, and the growth pattern of its community over the years.

The original work of preferential attachment model predicts a power-law degree distribution with exponent $3$, though empirically this value varies from $2$ to $3$ for most real world networks~\cite{barabasi1999mean, albert2002statistical}. By measuring this exponent we can see the fairness of the distribution of followers among users. Combined with the above-mentioned methods we can also identify whether old or new users have the most followers.

The growth pattern of the GitHub community includes 1) the average degree growth curves of users who join GitHub at different time periods, and 2) how the size of the community grows over the years. By comparing the average degree growth curves from different time periods we can see whether the growth favors old or new users. The growth of the network size also greatly influence the growth of node degrees~\cite{sun2020timeinvariant}.

\section{Experiments and Results}
\label{results}
In this section, we first introduce the datasets that we collect and use, and then present the experimental results based on the methodologies described in Section~\ref{methodology}.

\subsection{Datasets}
\label{dataset}
The experiments were carried in two separate datasets, as shown in Table \ref{table:data_collection}. The GHTorrent dataset~\cite{Gousi13} was used for the node centrality based analysis and the macroscopic analysis. The other dataset, which was crawled through GitHub API, was used for the user activity based analysis. Below is a detailed description of the two datasets.

\subsubsection{GHTorrent}
This dataset contains historical data of GitHub from 2007-10-20 to 2018-09-29, including all user profiles and activities such as following, project, pull request and issue history. After preprocessing, we filter out 1) the accounts with no more than 5 followers\footnote{The average in-degree of the network is approximately 5.}, 2) the organizational accounts, and 3) fake or invalid accounts\footnote{https://ghtorrent.org/relational.html} for some analyses. Furthermore, we build the follower network of GitHub on top of the GHTorrent data, in which each node represents a user, and each directed link from one node to another represents a ``follow'' relation. Table~\ref{table:ghtorrentstats} shows the statistics of the GHTorrent dataset.

\begin{table}[H]
\caption{Statistics of the GHTorrent dataset}
\label{table:ghtorrentstats}
\centering
\begin{tabular}{c|c|c}
\toprule
                   & \textbf{\# Users} & \textbf{\# Links} \\ \midrule
Original dataset   & 5,375,944         & 29,809,738 \\ \midrule
\makecell{After filtering fake, deleted\\ and organization users}   & 5,134,221 & 27,874,082 \\ \midrule
\makecell{After filtering users \\ with $\leq$ 5 followers}         &   760,467 & 16,546,310 \\ \bottomrule
\end{tabular}
\end{table}

\subsubsection{Github API}
We also collect data by using the GitHub REST API\footnote{https://developer.github.com/v3}. GitHub has restrictions on the number of API requests that can be made to their servers to ensure proportionate performance. It only allows 50 unauthenticated API requests per hour and when that limit is reached, the API no longer shows results. To overcome this challenge we use authenticated API requests to increase the rate limit from 50 to 5000 per hour. The subsequent response to API request resulted in JSON data to be parsed for further analyses.

The features and datasets used in each analysis are shown in Table~\ref{table:data_collection}.

\begin{table}[H]
\caption{Features and datasets for each methodology}
\centering
\label{table:data_collection}
\begin{tabular}{@{}c|c|c@{}}
\toprule
Analysis & Features & Dataset\\ \midrule
Node centrality based   & \makecell{Followers \\ HITS \\ PageRank} & GHTorrent \\ \midrule
User activity based     & \makecell{Pull requests \\ Issues} & Github API \\ \midrule
Macroscopic             & Followers & GHTorrent \\ \bottomrule
\end{tabular}
\end{table}

\subsection{Node centrality based analysis}

\begin{figure}[H]
\centering
\includegraphics[scale=0.60]{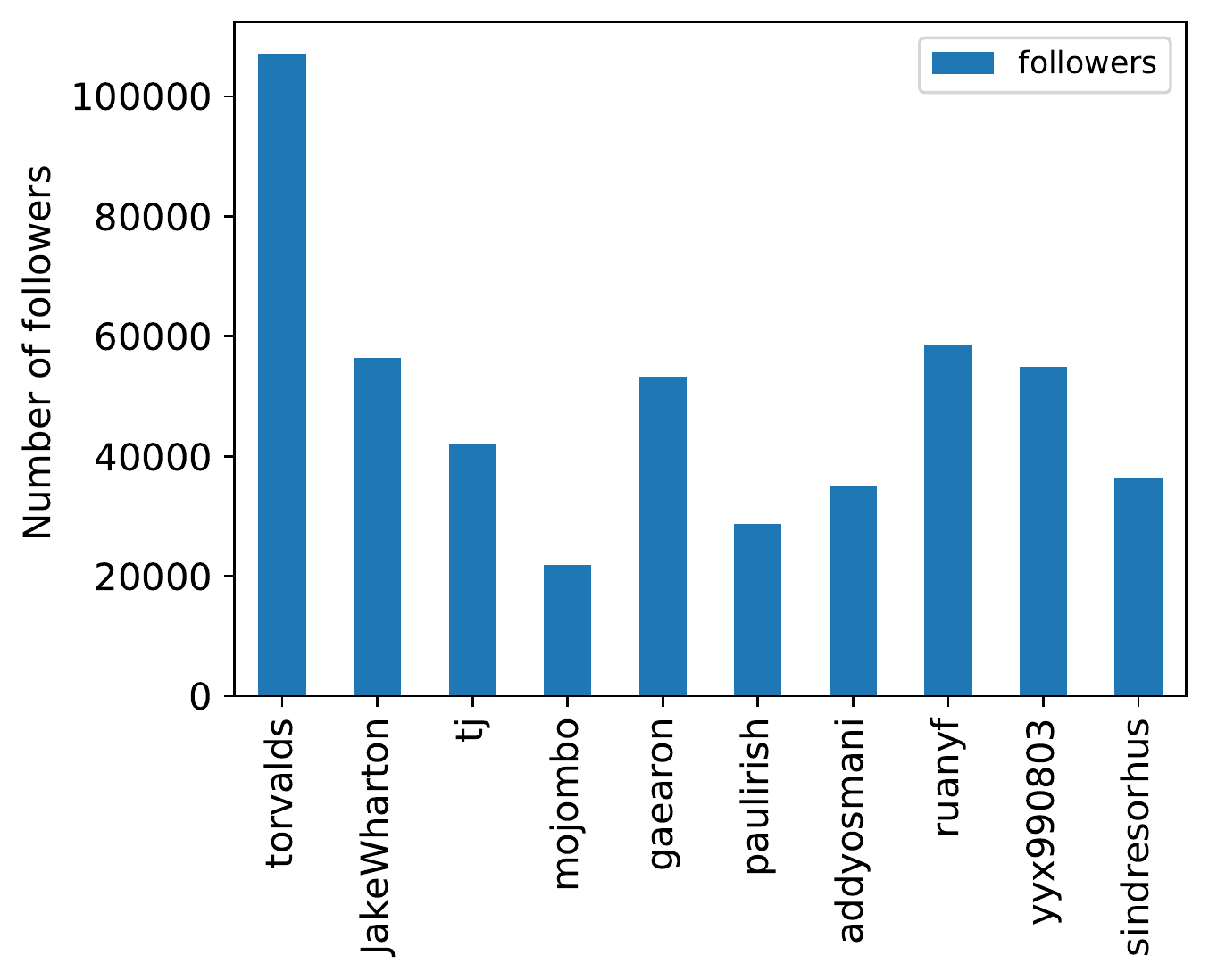}
\caption{The follower numbers of the users with top 10 PageRank.}
\label{fig:highfol}
\end{figure}

The top users with the highest number of followers and PageRank are quite similar. This classification is a proven method to identify important users in a network \cite{ramalingam2018fake} while discarding those that have fake followers~\cite{mehrotra2016detection}. The top 10 influential users based on their PageRank and followers values are shown in Fig.~\ref{fig:highfol}.
The analysis of Fig.~\ref{fig:highfol} reveals that follower count depicts quantitative, but not qualitative influence. For instance, even though the user “tj” has comparatively less number of followers than user “ruanyf”, still the user “tj” gets top position due to reason of being followed by influential users in the network.

\begin{figure}[H]
\centering
\includegraphics[scale=0.60] {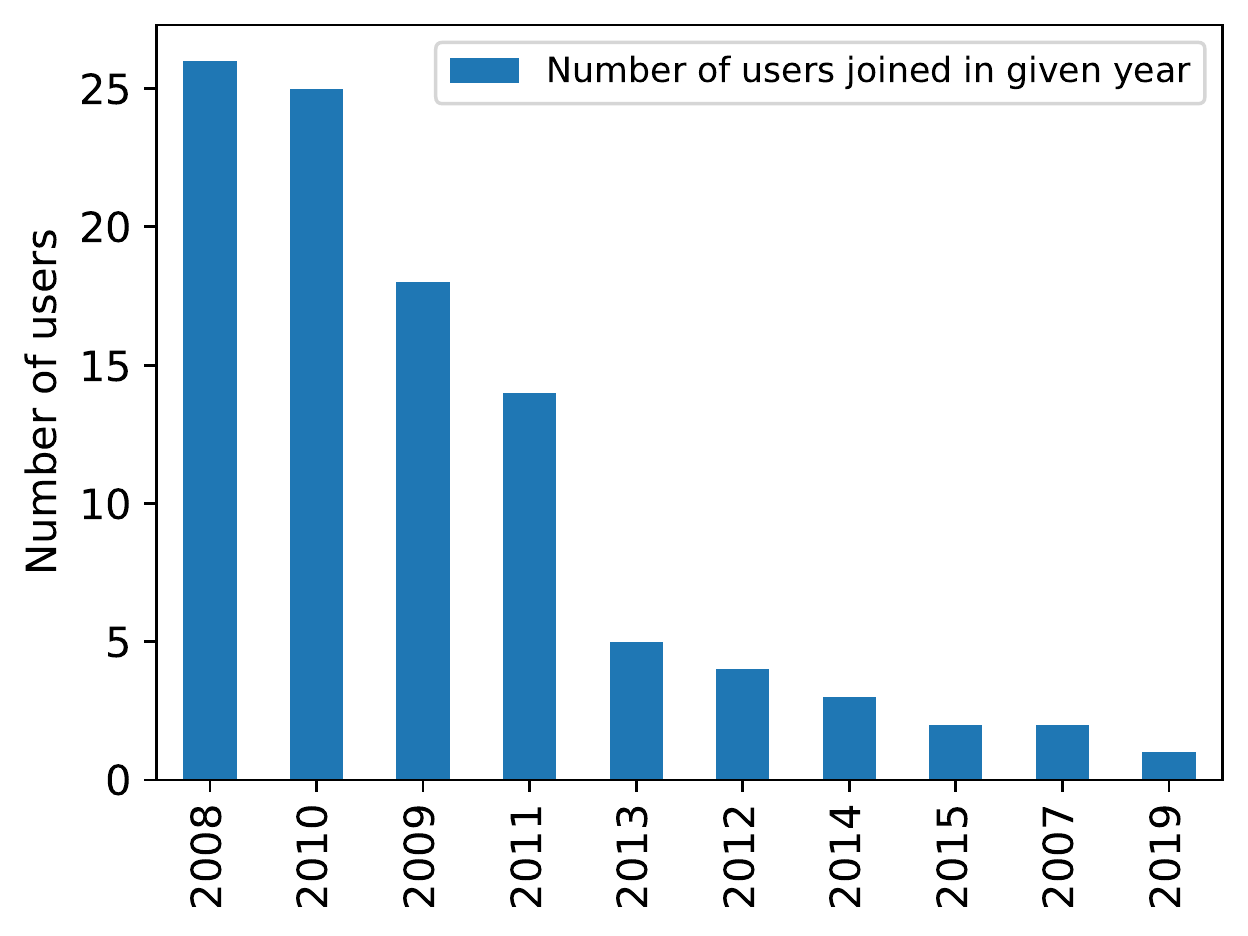}
\caption{Distribution of the joining years of the top 100 PageRank users.}
\label{fig:highfolyear}
\end{figure}

We also performed analysis for the top 100 users based on PageRank values against their joining date on GitHub. It was observed that, the year 2008 and 2010 witnessed as high as 26\% and 25\% of influential users joining the platform. Also, less users followed the top influential users after these years. In addition, the early year of joining GitHub does not have significant impact in garnering high number of followers. Fig. \ref{fig:highfolyear} further reveals that number of followers of influential users joining early is not monotonically increasing. This illustrates that early joining is not an important factor to be an influencer on GitHub.

\subsection {User activity based analysis}

\begin{figure}[H]
\centering\subfloat[\label{fig:top10vsmerge}]{\includegraphics[width=0.48\linewidth]{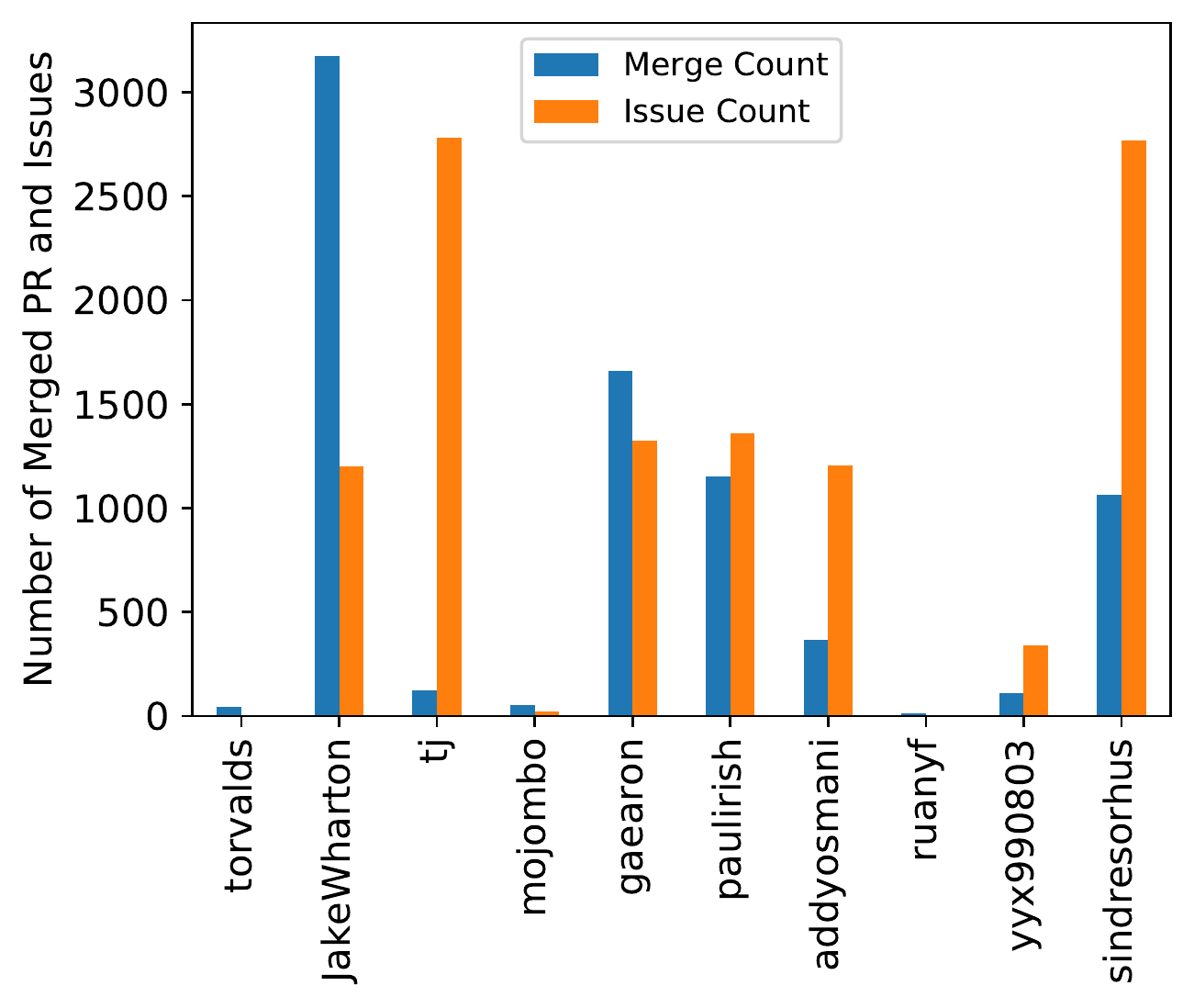}}\quad \quad\subfloat[\label{fig:top10bymerge}]{\includegraphics[width=0.48\linewidth]{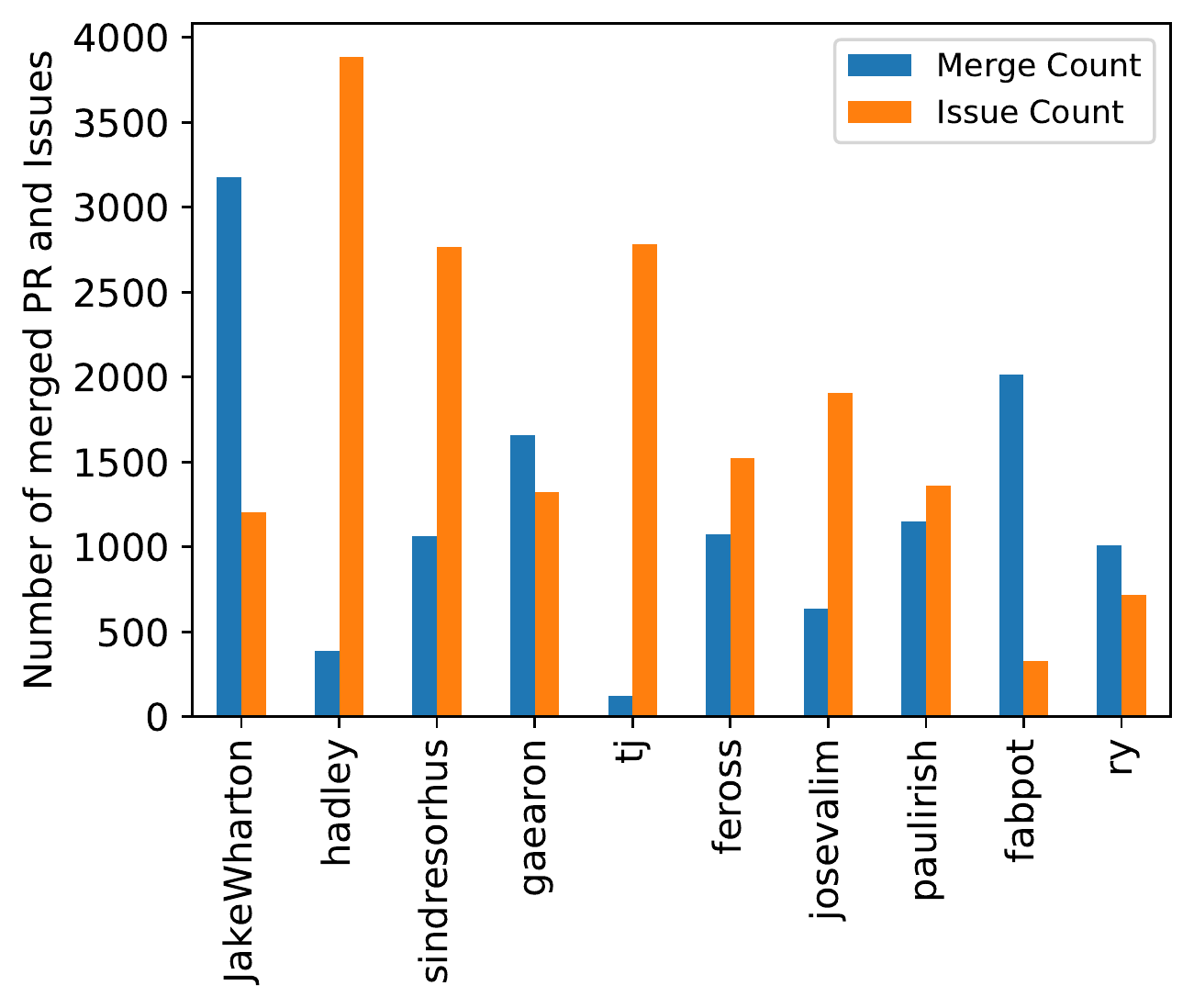}}\caption{Comparison of the merge counts and issue counts of the top 10 influential users in the software social network perspective, as measured by (a) PageRank, (b) merged PR and issue count.}\end{figure}

If we consider the PageRank of the network as the basis of user popularity and check the coding contribution made by such users, we find that equal share of such users have an insignificant contribution (Fig.\ref{fig:top10vsmerge}). Still, just due to the high values of their PageRank, they enjoy a high reputation in the network. On the other hand, if we consider the created issues and merged Pull Requests (PR) as metrics to identify influential users~\cite{goggins2014connecting}, we find that five new people out of ten are coming in the list based on their contribution to the software community. This is an interesting insight that contribution is a significant determinant for being an influencer apart from being popular.
From Fig. \ref{fig:top10vsmerge} and \ref{fig:top10bymerge}, it is evident that the user “Torvalds”, who enjoyed the top position in previous ranking disappears from this list of top 10 contributors. Also, another user “JakeWharton”, who possessed second position in previous ranking rises to the top position in this list. This reveals a fascinating insight that the user “Torvalds” is an influential user, based on only his high number of followers, but the user “JakeWharton” is definitely an influential user based on the high number of followers as well as based on his contributions. From the perspective of software contributions, the latter ranking metrics is more promising and conclusive to identify an influential user in a network.
Another metric, the merge ratio, shows the acceptance of a user’s coding contribution in the network. However, since it calculates the accepted pull requests as a proportion of submitted pull requests, this can be biased metric due to the reason that a user can submit a smaller number of requests, the majority of which gets approved and user can be defined an influential. However, if another user makes significant contribution by submitting a high number of requests, and due to the context of the problem, a coding contribution can be rejected even though it was significant. We do not consider the merge ratio for analysis due to this reason. \par
This indicates to the result that some users are highly popular just based on their high follower count and usually contribute quite less, however other users get to contribute a lot and thus, gain the follower count, and so have high follower count as well as high contribution.

\subsection{Correlation among  metrics}
Based on network and software contribution metrics for the users with top 100 PageRank, we calculated the Pearson correlation values $\rho$ among different metrics, as shown in Table \ref{table:corr_coeff}. We find that ``Issue Count'' metric is slightly correlated with ``Merged PR Count'' measure, with a correlation value of 0.427. This suggests that if the number of merged pull requests by a user is high, there are more chances of the same user contributing by creating and working on the issues. This correlation matrix helped us to narrow down the approach of analysis based on pull requests and issues.
\begin{table}[!htb]
\caption{Correlation coefficients among parameters}
\centering
\label{table:corr_coeff}
\begin{tabular}{@{}c|r|r|r|r@{}}
\toprule
$\rho$ & Followers & Repositories & Merged PR & Issues \\ \midrule
Followers & 1 &  &  &  \\ \midrule
Repositories & -0.03809 & 1 &  &  \\ \midrule
Merged PR & 0.271839 & 0.00603 & 1 &  \\ \midrule
Issues & 0.23514 & 0.05876 & 0.4271 & 1 \\ \bottomrule
\end{tabular}
\end{table}

\subsection{Macroscopic analysis}
In this section we present our analysis based on the macroscopic observations of the follower network of GitHub.

\begin{figure}[H]
\centering
\includegraphics[scale=0.60]{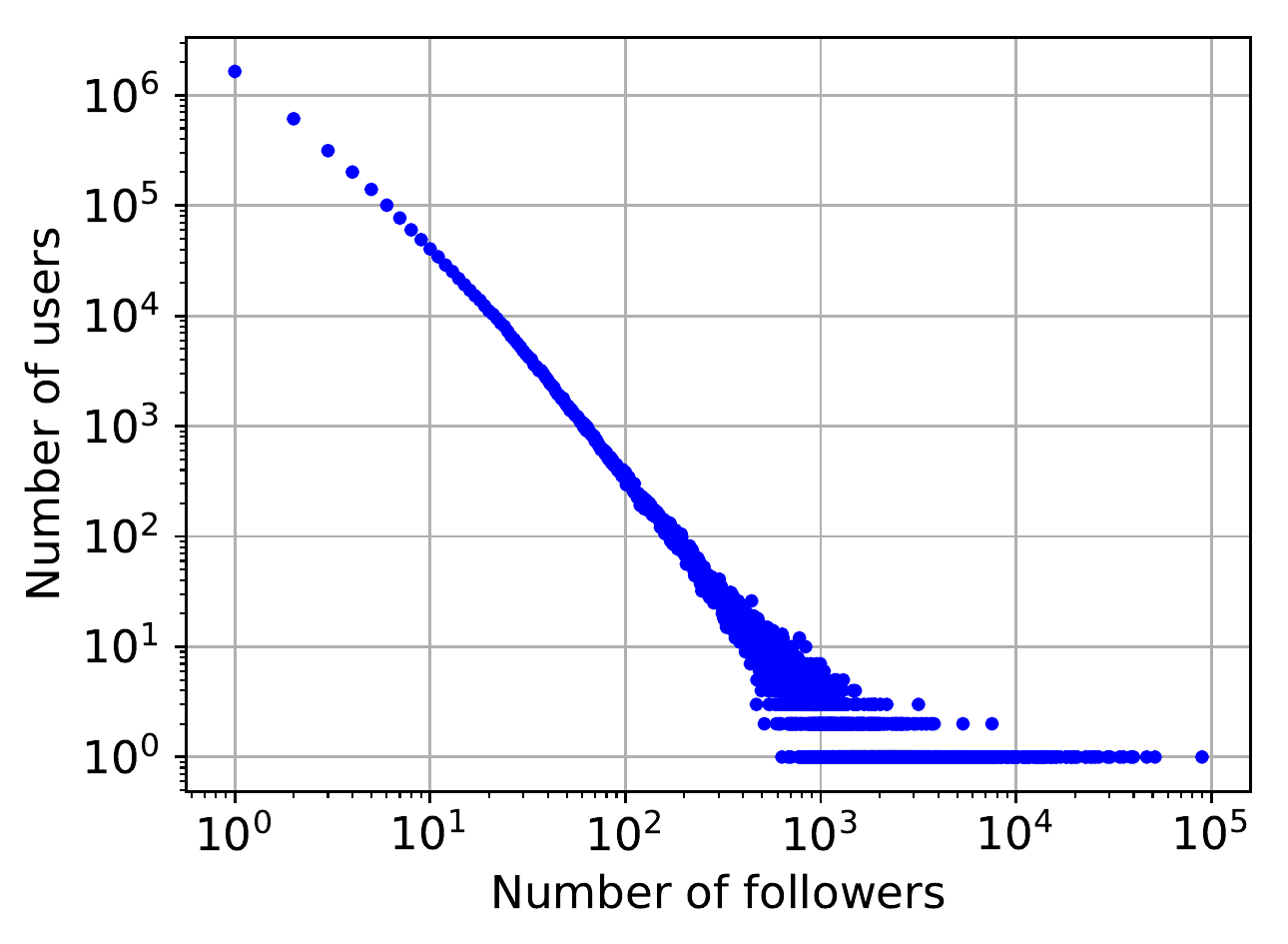}
\caption{The in-degree distribution of GitHub users. More than five million users have 100 or less followers. In comparison, only 61 users have more than 10,000 followers. Note: this plot uses the log-log scale.}
\label{fig:degdistr}
\end{figure}

Fig.~\ref{fig:degdistr} shows the in-degree (i.e., the number of followers) distribution of GitHub users. The nearly linear curve (especially in the domain of low in-degrees) with a long tail in the log-log plot indicates that the distribution follows power-law approximately, and the network has a scale-free structure: The majority of users have only few followers, whilst a small fraction of users have extremely large numbers of followers, up to the magnitude of $10^5$.

Combined with the analysis shown in the previous subsections, this can be an indication of the presence of the first mover advantage in the GitHub network, as the few users with large numbers of followers are the ones who have benefited from the rich-gets-richer effect over time. However, possibly due to aging, no evidence of the winner-takes-all effect is present yet.

\begin{figure}[H]
  \centering
  \subfloat[\label{fig:avgfol}]
    {\includegraphics[width=0.48\linewidth]{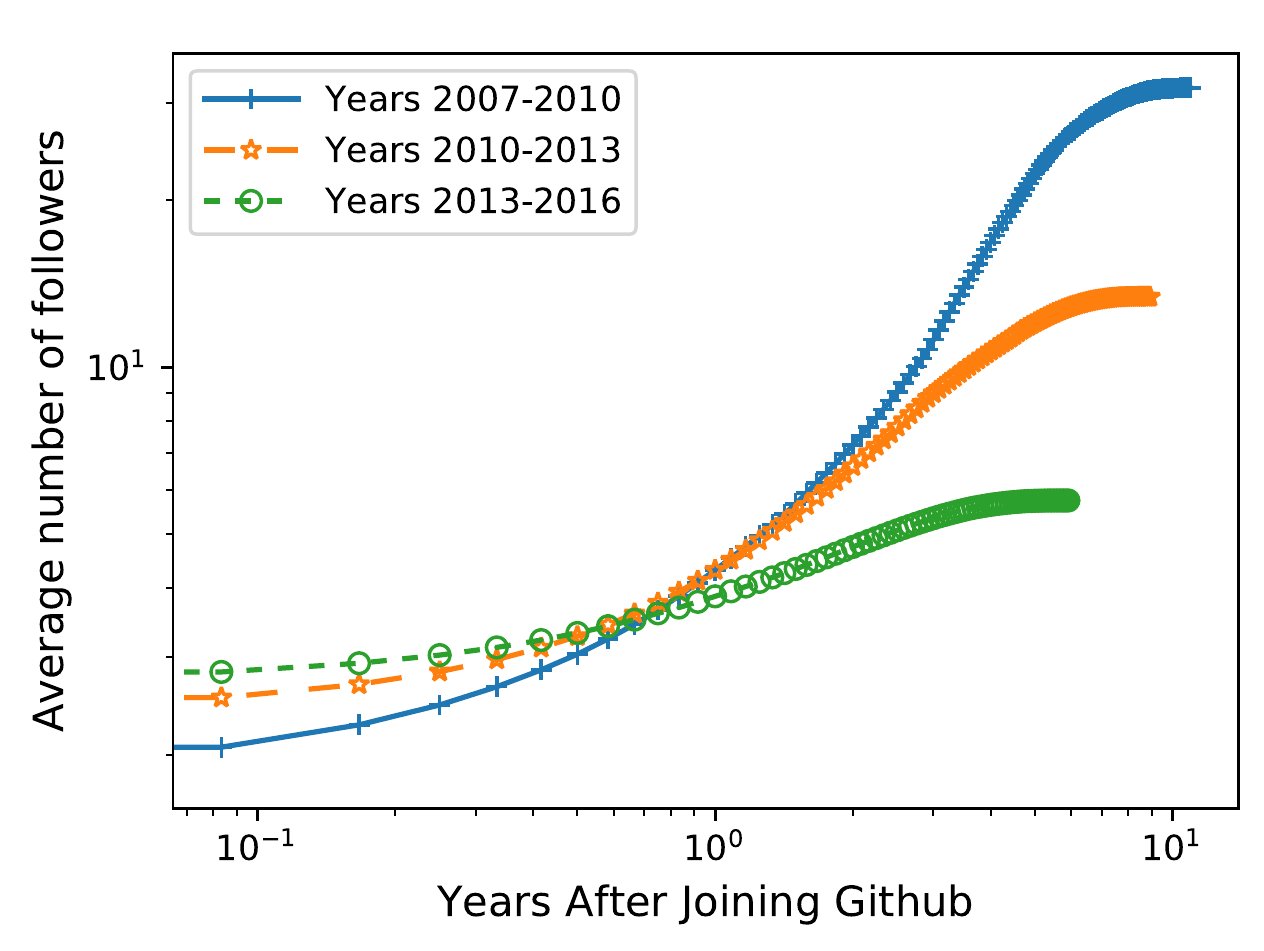}}
    \quad \quad
  \subfloat[\label{fig:networkgrowth}]
    {\includegraphics[width=0.48\linewidth]{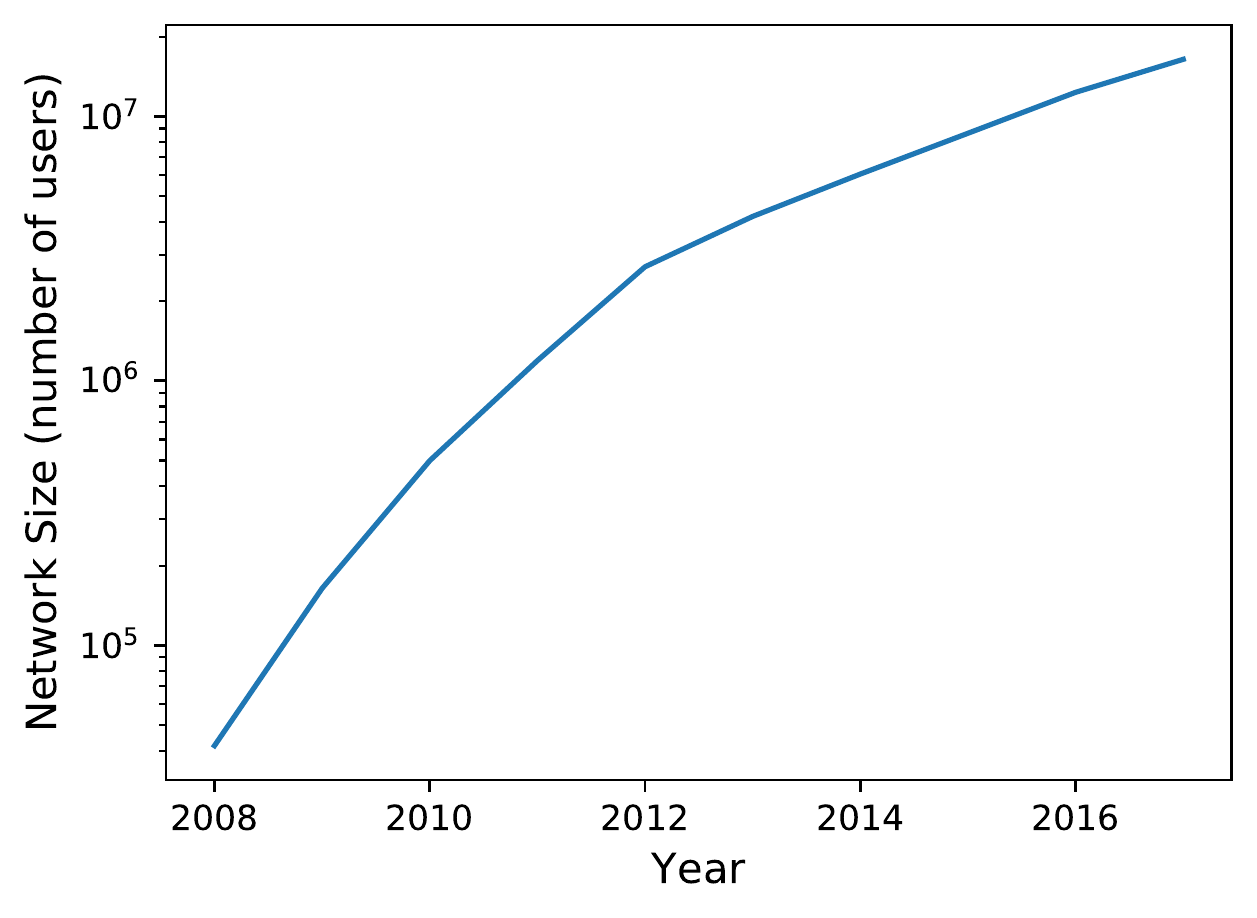}}
    \caption{Growth patterns of the GitHub network over the years. Panel (a) shows the average number of followers as a function of user's lifetime in GitHub the log-log scale. Users are grouped by the year they joined GitHub. Panel (b) shows the growth of GitHub user numbers over the years in the linear-log scale.}
\end{figure}

In Fig.~\ref{fig:avgfol} we group the users according to the years they joined GitHub, and plot the average in-degree as a function of their GitHub lifetime
separately for users who joined in different periods of time. Users with no followers as well as users who joined later than 2016 are excluded, as they hardly contribute to meaningful observations.
As we can see from the plot, given enough period of time (longer than one year) to grow, with the same GitHub lifetime to give a fair comparison, the users who joined earlier have a larger average in-degree than the ones who joined later, despite of the fact that they started with the right opposite situation. This is different to what has been observed from academic citation networks where the average degree growth is time invariant~\cite{sun2018decay, sun2020timeinvariant}, and is a strong indication of the first mover advantage.

We also plot the growth curve of the total number of users on GitHub from Year 2008 to 2019 in Fig.~\ref{fig:networkgrowth}. The concave curve in the linear-log plot indicates that the total user number underwent a sub-exponential growth over the years. Although this does not mean the absolute number of new users is declining each year, the growth is still slower than exponential to sustain a time-invariant degree growth which is a required property of the ``no first mover advantage'' case, according to analytical models~\cite{sun2020timeinvariant}. Hence, in GitHub, the preferential attachment process favors old users who have accumulated influences for a longer time, and effectively results in the first mover advantage.

\subsection{Comparison with other studies}
To consider evaluation and validation of results, we compare our results with existing research works.    
To identify influential users, analysis was done by Yan et al \cite{hu2018user} using dataset of all Github users between April 1, 2008 and January 19, 2017 as fetched from GHtorrent database dump. We find significantly similar results for top influential users list based on followers, HITS and PageRank algorithm. Top users are similar, just the ranking of our results change from their results. Similarly, we find extensive similarity in results by comparison with list of influential users from studies done by Bana et al \cite{bana2018influence}. However, their data set was crawled through Google BigQuery as well as GitHub API and the approach consisted of creating a developer-developer network, in which a connection is created based on the fact if two developers worked together in any projects on GitHub. This similarity in top results can be attributed to the effect of first mover advantage because they could hold influential position in the network. Different users were ranked in different position when evaluated against different metrics (see appendix \ref{appendix:top10}).

\section{Conclusion}\label{conclusion}
In this paper, we present the influential behaviors of the users in the GitHub network leveraging node centrality based and user activity based approaches. We observed that early adopter dynamics exist in users who has high number of followers. The metrics to measure influence are the number of followers which helps to obtain high PageRank values, pull requests and issues which reflect contribution of users. From the macroscopic observation, we observe that early adopters of GitHub are found out to be having higher follower users and with the growth in network size, this phenomenon was grown as well helping them get an influential position in the network. From the perspective of network characteristics as well as the software contribution perspective, few users are just having a large number of followers but rarely contribute actively. On the other hand, many users have less number of followers, however, are quite active in the community as depicted by their contribution in coding activities. 

As future work, we plan to add controversy features such as sentiment, stances of issue discussions as user activity metrics. Such and other factors determine the quality of information, content, and products on the social network deliberation. Further study on the dynamics of democratic consensus-finding has the potential to strengthen collective resilience against threats to degrade information, content, and products. Thus, this stream of research relates deeply to Web governance and even offline life.

\begin{acksname}
Baris and Hosseini work for the EU Horizon 2020 project Co-Inform (\url{www.coinform.eu}) under contract No.~770302.
Sun works for the EU Horizon 2020 project CUTLER (\url{www.cutler-h2020.eu}) under contract No.~770469.
\end{acksname}


\bibliography{references}
\bibliographystyle{unsrt}  \newpage
\begin{appendix}
\section{Users ranked on the basis of different metrics}\label{appendix:top10}
\begin{table*}[ht]
\centering
\label{table:summ}
\begin{tabular}{>{\centering\arraybackslash}m{1cm}|>{\centering\arraybackslash}m{2cm}|>{\centering\arraybackslash}m{2cm}|>{\centering\arraybackslash}m{2cm}|>{\centering\arraybackslash}m{2cm}|>{\centering\arraybackslash}m{2cm}}
\toprule
\textbf{User Rank} & \textbf{PageRank} & \textbf{Authority Rank (HITS)} & \textbf{Followers count} & \textbf{Merged Pull Requests} & \textbf{Number of Issues} \\
\midrule
1 & \makecell{0.00816525 \\ torvalds} & \makecell{0.00004355 \\ torvalds} & \makecell{105277 \\ torvalds} & \makecell{3175 \\ JakeWharton} & \makecell{3887 \\ hadley} \\ \midrule
2 & \makecell{0.00242481 \\ JakeWharton} & \makecell{0.00004298 \\ tj} & \makecell{58191 \\ ruanyf} & \makecell{2016 \\ fabpot} & \makecell{2782 \\ tj} \\ \midrule
3 & \makecell{0.00215897 \\ tj} & \makecell{0.00003689 \\ addyosmani} & \makecell{56035 \\ JakeWharton} & \makecell{1661 \\ gaearon} & \makecell{2769 \\ sindresorhus} \\ \midrule
4 & \makecell{0.00174284 \\ mojombo} & \makecell{0.00003660 \\ sindresorhus} & \makecell{54043 \\ yyx990803} & \makecell{1153 \\ paulirish} & \makecell{1905 \\ josevalim} \\ \midrule
5 & \makecell{0.00152582 \\ gaearon} & \makecell{0.00003521 \\ paulirish} & \makecell{51842 \\gaearon} & \makecell{1075 \\ feross} & \makecell{1523\\ feross} \\ \midrule
6 & \makecell{0.00143539 \\paulirish} & \makecell{0.00003092 \\yyx990803} & \makecell{41876 \\tj} & \makecell{1062 \\sindresorhus} & \makecell{1361 \\paulirish} \\ \midrule
7 & \makecell{0.00139387 \\addyosmani} & \makecell{0.00002868 \\gaearon} & \makecell{36142 \\sindresorhus} & \makecell{1008 \\ry} & \makecell{1324 \\gaearon} \\ \midrule
8 & \makecell{0.00138717 \\ruanyf} & \makecell{0.00002776 \\defunkt} & \makecell{36002 \\llSourcell} & \makecell{812 \\mitsuhiko} & \makecell{1307 \\mbostock} \\ \midrule
9 & \makecell{0.00117444 \\yyx990803} & \makecell{0.00002750 \\substack} & \makecell{34824 \\addyosmani} & \makecell{733 \\mitchellh} & \makecell{1212 \\shiffman} \\ \midrule
10 & \makecell{0.00116498 \\sindresorhus} & \makecell{0.00002683 \\kennethreitz} & \makecell{29950 \\michaelliao} & \makecell{660 \\taylorotwell} & \makecell{1207 \\addyosmani} \\ \bottomrule
\end{tabular}
\end{table*}
\end{appendix}

\end{document}